\documentclass[a4paper,useAMS,usenatbib]{mnras}
\usepackage{graphics,epsfig,psfig}
\hypersetup{colorlinks=true,allcolors=blue}
\usepackage{todonotes}
\usepackage[normalem]{ulem}
\usepackage{xcolor}
\usepackage{float}
\usepackage{booktabs}
\usepackage[style=base, singlelinecheck=off, font=small, labelfont=bf, 
justification=raggedright]{caption}
\usepackage{subfigure}

\usepackage{graphicx}
\usepackage[]{inputenc,amssymb}
\usepackage{natbib}
\usepackage{url}
\usepackage{widetext}
\usepackage{amsmath}
\usepackage{cancel}

  \makeatletter
\def\@fnsymbol#1{\ensuremath{\ifcase#1\or \dagger\or \ddagger\or
   \mathsection\or \mathparagraph\or \|\or **\or \dagger\dagger
   \or \ddagger\ddagger \else\@ctrerr\fi}}
    \makeatother
\def \be{\begin{equation}}
\def \ee{\end{equation}}
\def \bea{\begin{eqnarray}}
\def \eea{\end{eqnarray}}

\def\apj{ApJ}
\def\mnras{MNRAS}

\def\aap{A\&A}
\definecolor{webgreen}{rgb}{0,.5,0}
\definecolor{webbrown}{rgb}{.6,0,0}

\def\aap{A\&A}
\def\apj{ApJ}
\def\apjs{ApJS}

\def\mnras{MNRAS}

\def\aj{AJ}

\def\jcap{JCAP}
\def\sovast{Soviet Astronomy}

\title[Sub-solar mass PBHs]{Prospects of discovering  sub-solar primordial black holes using the stochastic gravitational wave background from  third-generation detectors}
\author[Mukherjee, Meinema \& Silk (2021)]{Suvodip Mukherjee$^{1,2,3,4}$\thanks{s.mukherjee@uva.nl},  Matthew S. P. Meinema$^{1}$ \thanks{mattmeinema@hotmail.com}, Joseph Silk$^{5, 6,7} $\thanks{silk@iap.fr}\\
$^{1}$ Gravitation Astroparticle Physics Amsterdam (GRAPPA),
Anton Pannekoek Institute for Astronomy and Institute for Physics,\\
University of Amsterdam, Science Park 904, 1090 GL Amsterdam, The Netherlands\\
$^2$ Institute Lorentz, Leiden University, PO Box 9506, Leiden 2300 RA, The Netherlands\\
$^3$Delta Institute for Theoretical Physics, Science Park 904, 1090 GL Amsterdam, The Netherlands\\
$^4$Perimeter Institute for Theoretical Physics, 31 Caroline Street N., Waterloo, Ontario, N2L 2Y5, Canada\\
$^{5}$ Institut d'Astrophysique de Paris, UMR 7095, CNRS, Sorbonne Universit\'e, 98bis Boulevard Arago, 75014 Paris, France\\
$^{6}$ The Johns Hopkins University, Department of Physics \& Astronomy, 3400 N. Charles Street, Baltimore, MD 21218, USA\\
$^{7}$ Beecroft Institute for Cosmology and Particle Astrophysics, University of Oxford, Keble Road, Oxford OX1 3RH, UK\\
}
\pubyear{2021}
\begin{document}
\label{firstpage}
\pagerange{\pageref{firstpage}--\pageref{lastpage}}
\maketitle

\label{firstpage}

\begin{abstract}
Primordial black holes (PBHs) are dark matter candidates that  span broad mass ranges from $10^{-17}$ $M_\odot$ to $\sim 100$ $M_\odot$. We show that the stochastic gravitational wave background can be a powerful window for the detection of sub-solar mass PBHs and shed light on their formation channel via third-generation gravitational wave detectors such as Cosmic Explorer and the Einstein Telescope. By using the mass distribution of the compact objects and the redshift evolution of the merger rates, we can distinguish astrophysical sources from  PBHs and will be able to constrain the fraction of sub-solar mass PBHs $\leq 1$ $M_\odot$ in the form of dark matter $f_{PBH}\leq 1\%$  at $68\%$ C.L. even for a pessimistic value of a binary suppression factor.
In the absence of any suppression of the merger rate, constraints on $f_{PBH}$ will be less than $0.001\%$. Furthermore, we will be able to measure the redshift evolution of the PBH merger rate with about $1\%$ accuracy, making it possible to uniquely distinguish between the Poisson and clustered PBH scenarios. 
\end{abstract}

\section{Introduction}
The existence of black holes (BHs) below the Chandrasekhar mass limit $1.4$ M$_\odot$ \citep{1931ApJ....74...81C} offers a distinct signature to unveil the existence of alternative channels to form BHs \citep{1967SvA....10..602Z,1971MNRAS.152...75H,1975ApJ...201....1C, 1980PhLB...97..383K,Dasgupta:2020mqg}, apart from those of astrophysical origin. One of the promising channels of formation of BHs for masses lighter than $1.4$ M$_\odot$, is through  formation in the early Universe \citep{1967SvA....10..602Z,1971MNRAS.152...75H,1975ApJ...201....1C, 1980PhLB...97..383K, 1985MNRAS.215..575K,Carr:2005zd, Clesse:2016vqa, Sasaki:2018dmp, Sasaki:2016jop, Raidal:2017mfl, Raidal:2018bbj, Vaskonen:2019jpv, Gow:2019pok,Jedamzik:2020ypm,Jedamzik:2020omx,DeLuca:2020jug, Atal:2020igj, DeLuca:2020qqa}. {This issue has been revitalized by events such as  GW190814 \citep{LIGOScientific:2020zkf, Vattis:2020iuz} that features a BH merger with a $2.5-2.67$ M$_\odot$ compact object that is outside the usual neutron star (NS)  mass range. NS-BH mergers have been recently detected, with NS masses below the NS maximal mass limit of  $\sim 2 \rm M_\odot$ \citep{LIGOScientific:2021qlt}. To bridge this mass gap for the GW190814 event, one may have recourse to a massive NS that underwent collapse to a BH eg, \citep{Most:2020bba}, or else appeal to a primordial black hole (PBH). In the former case, there should be a hitherto undetected \citep{GravityCollective:2021kyg, deWet:2021qdx, Alexander:2021twj}  electromagnetic signal,  and in the latter case, one would expect a companion population of subsolar mass  PBHs. The latter is well-motivated theoretically by the generation of large fluctuations at the QCD phase transition \citep{Clesse:2020ghq}. This paper will explore how one may detect such a population of subsolar PBHs via the stochastic gravitational wave background (SGWB).
}

The existence of PBHs can shed light on the physics of the early Universe and can also provide a candidate for the dark matter which makes up about $23\%$ of the mass-energy budget of the Universe \citep{Spergel:2003cb, 2011ApJS..192...18K, Ade:2013zuv, Anderson:2013zyy, Cuesta:2015mqa, Ade:2015xua, Aghanim:2018eyx, Alam:2016hwk}. However, to discover PBHs and identify them as the dark matter candidates, we need to understand the abundance of these sources and also distinguish them from compact objects of astrophysical origin. Two key features which can be used to distinguish between astrophysical black holes (ABHs) and PBHs are the mass and redshift distributions. As ABHs cannot be smaller than $\sim 2$ M$_\odot$ and would not form before the birth of stars, one would expect the mergers of compact objects of astrophysical origin to follow the cosmic star formation rate (SFR) history of the Universe \citep{Madau:2014bja} and will not have a peak in the merger rate at a redshift higher the SFR peak,  around $z\sim 2$ according to current multi-frequency electromagnetic observations \citep{Madau:2014bja}. 

One of the promising windows for detecting such sub-solar BHs is using the gravitational wave signal emitted during their coalescence. From the ongoing network of gravitational wave detectors such as advanced LIGO \citep{TheLIGOScientific:2014jea} and Virgo \citep{TheVirgo:2014hva}, bounds on the local Universe BH merger rate have been obtained \citep{Abbott:2018oah, Authors:2019qbw}, and the possibility of sub-solar BHs constituting a significant fraction of dark matter remains a possibility \citep{Phukon:2021cus}. However, due to the weak strength of the gravitational wave signal from sub-solar masses, these sources can only be confidently detected
(i.e., with a matched filtering \citep{Sathyaprakash:1991mt, Balasubramanian:1995bm}  SNR above eight) as individual events
at very low redshift with the current network of GW detectors, $z<0.1$. As a result, the redshift evolution of the mergers, which is one of the distinct features to identify PBHs as dark matter cannot be explored currently for sub-solar masses.   
 
However, one of the promising windows for probing high redshift mergers of the GW sources is through the SGWB which arises from unresolved coalescences of binary systems. Although the SGWB for a single source is going to be subthreshold, it can nevertheless be detected by integrating over a large observation time and combining the contributions from multiple unresolved sources. In this paper, we estimate the SGWB signal from sub-solar mass compact objects, and the possibility of detection via upcoming gravitational wave detectors. By exploring the evolution of the merger rate of the sub-solar sources and their mass distribution, we will be able to distinguish between sources of primordial origin and astrophysical origin as proposed by \citep{Mukherjee:2021ags}. Along with the SGWB, individual events can also shed light on the existence of subsolar PBHs \citep{Abbott:2018oah, Authors:2019qbw, Chen:2019irf, Phukon:2021cus}.


\section{The formalism of Stochastic GW background}
The SGWB from the coalescence of binary systems arise from the sources which cannot be detected as individual events for a given network of GW detectors. These sources contribute to a background energy density which can be written with respect to the critical energy density $\rho_c c^2$ of the Universe in terms of a dimensionless parameter $\Omega_{GW}(f)$ as  \citep{Allen:1996vm,Phinney:2001di}
\begin{align}\label{sgwb-1}
    \begin{split}
        \Omega_{GW} (f)= &\frac{1}{\rho_cc^2} \int d\theta  \int^\infty_{z_{\text{min}} (\theta)} dz    \overbrace{\frac{dV}{dz}}^{\text{cosmology}}\overbrace{\frac{p(\theta)\mathcal{R}_{GW}(z, \theta)}{(1+z)}}^{\text{astrophysics}} \\& \times \overbrace{\bigg(\frac{1+z}{4\pi c d_L^2}\frac{(G\pi)^{2/3}\mathcal{M}_c^{5/3}f_r}{3} \mathcal{G}(f_r)\bigg)}^{\text{GW source}}\bigg|_{f_r= (1+z)f}, 
    \end{split}
\end{align}
where the luminosity distance to the sources is denoted by $d_L$ and the energy emission per frequency bin in the source frame is expressed in terms of the chirp mass $\mathcal{M}_c= \eta^{3/5}M$\footnote{We show the total mass $M= m_1 + m_2$ and the symmetric mass ratio $\eta= m_1m_2/M^2$.} of the gravitational wave sources, $\mathcal{R}_{GW}(z, \theta)$ is the merger rate of GW sources with source parameters $\theta$ having a probability distribution $p(\theta)$, and the function $\mathcal{G}(f_r)$, which captures the frequency dependence during the inspiral, merger, and ringdown phases of the gravitational wave signal, is  given by \citep{Ajith:2007kx}. 
The value of the lower-limit of the redshift integration  $z_{\text{min}} (\theta)$ depends on the source properties $\theta$ which contribute to the background. For a given network of GW detectors, the choice of $z_{\text{min}} (\theta)$ can be decided on the basis of the maximum redshift value up to which a GW source can be detected as an individual event using the matched-filtering technique $z^{\text{mf}}_{\text{max}}$. For the  ongoing network of GW detectors, the value of $z^{\text{mf}}_{\text{max}}<1$, and its contribution to the SGWB, are calculated for the combination of GW sources of both astrophysical and primordial origin  \citep{Allen:1996vm,Allen:1997ad, Phinney:2001di, Wu:2011ac, Regimbau:2007ed, Romano:2016dpx, Rosado:2011kv, Zhu:2011bd, Wang:2016ana, Mandic:2016lcn, 10.1093/mnras/stz3226, Callister:2020arv, Mukherjee:2021ags}.

\begin{figure}
      \includegraphics[trim={0.cm 0.cm 2.0cm 0.0cm},clip,width = 0.9\linewidth]{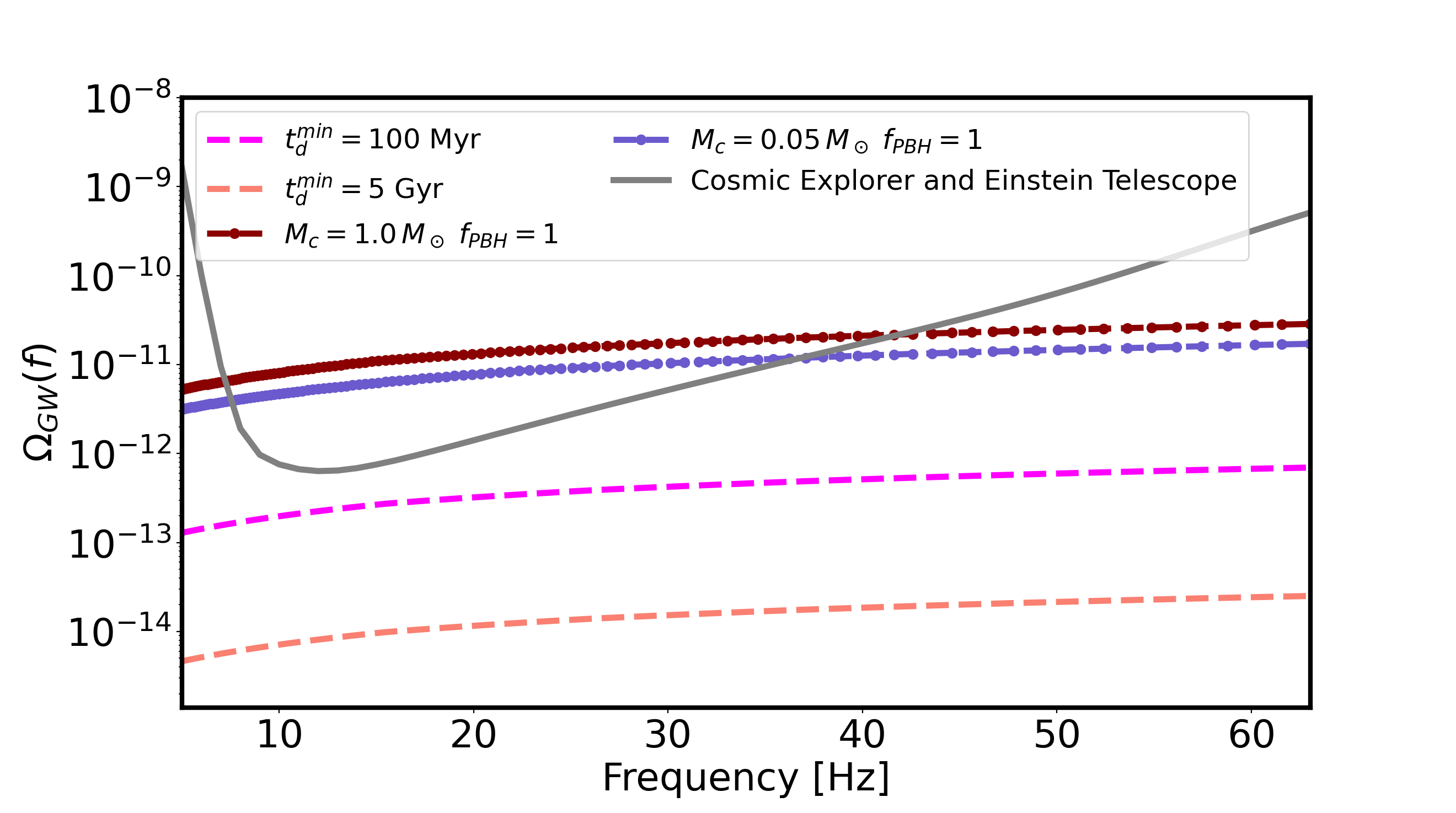}
    \caption{The SGWB signal for different characteristic mass-scales $M_c$ of PBH with $f_{PBH}=1$ and $\alpha=1.28$, along with the ABH cases for different values of the time-delay parameter $t_d^{min}$. The raw noise curve is shown in grey.}
    \label{fig:sgwb}
\end{figure}

However, for the third generation network of GW detectors such as Cosmic Explorer (CE) \citep{Evans:2016mbw} and Einstein Telescope (ET) \citep{Regimbau:2012ir, Maggiore:2019uih}, the maximum redshift up to which one can detect individual events is higher than for sources of individual masses above $1\, M_\odot$. GW sources with total masses in the source frame above $M=3\, M_\odot$ can be detected with a matched filtering signal-to-noise ratio $\rho_{th}>8$ up to redshift $z \sim 7$ \citep{2019CQGra..36v5002H}. So the astrophysical GW sources, which are expected to follow the Madau-Dickinson cosmic star-formation rate history \citep{Madau:2014bja}  are going to have a peak of the merger rate at redshift $z\leq2$ \citep{2010ApJ...716..615O,2010MNRAS.402..371B, 2012ApJ...759...52D, Dominik:2014yma, 2016MNRAS.458.2634M, Lamberts:2016txh, 2018MNRAS.474.4997C, Elbert:2017sbr, Eldridge:2018nop, Vitale:2018yhm, Buisson:2020hoq,Santoliquido:2020axb}. As a result, most of these sources are not going to contribute to the SGWB. However, if there exist compact objects of primordial origin, which are going to have a different merger rate from the ABHs, then such sources can be distinguished from astrophysical sources via the redshift evolution of the merger rate with the third generation detectors. We explain this in  Fig. \ref{fig:sgwb} by showing the detection horizon for different masses and how stochastic signals can distinguish the sources based on merger rates. The normalization of the curves is chosen such that both have the same value as at $z=0$. We also show the raw noise curve ([$\sum_{I, J, I\neq J}\frac{9H_0^4\gamma^2_{IJ}}{50\pi^4f^6S_{n_I}(f)S_{n_J}(f)}]^{-1/2}$) by the grey line for an integration time of five years  \footnote{The detector noise power spectrum for  detector $I$ (or $J$) is shown by $S_{n_I}(f)$ and the overlap reduction function by $\gamma_{IJ}$. The actual value of $\gamma_{IJ}$ for the third generation detectors will be known when the site and detector design are finalized.}. The possibility of distinguishing  ABHs from PBHs for the well-detected events is explored in an earlier study \citep{Chen:2019irf}. However, sources of masses below a total source frame mass $M \leq 2\, M_\odot$, can only be detected as individual events up to low redshift. As a result, if there exist PBHs of sub-solar masses, their mergers are going to be the main source of contributions to the SGWB, along with the contribution from binary neutron stars (BNSs). Two key features for distinguishing astrophysical sources from  PBHs are through their mass distributions and evolution of the merger rates with redshift. We discuss next these properties for the astrophysical compact object sources that can enable us to distinguish them from PBHs using the SGWB.

\begin{figure}
      \includegraphics[trim={1.cm 0.0cm 2.0cm 0.0cm},clip,width = 0.9\linewidth]{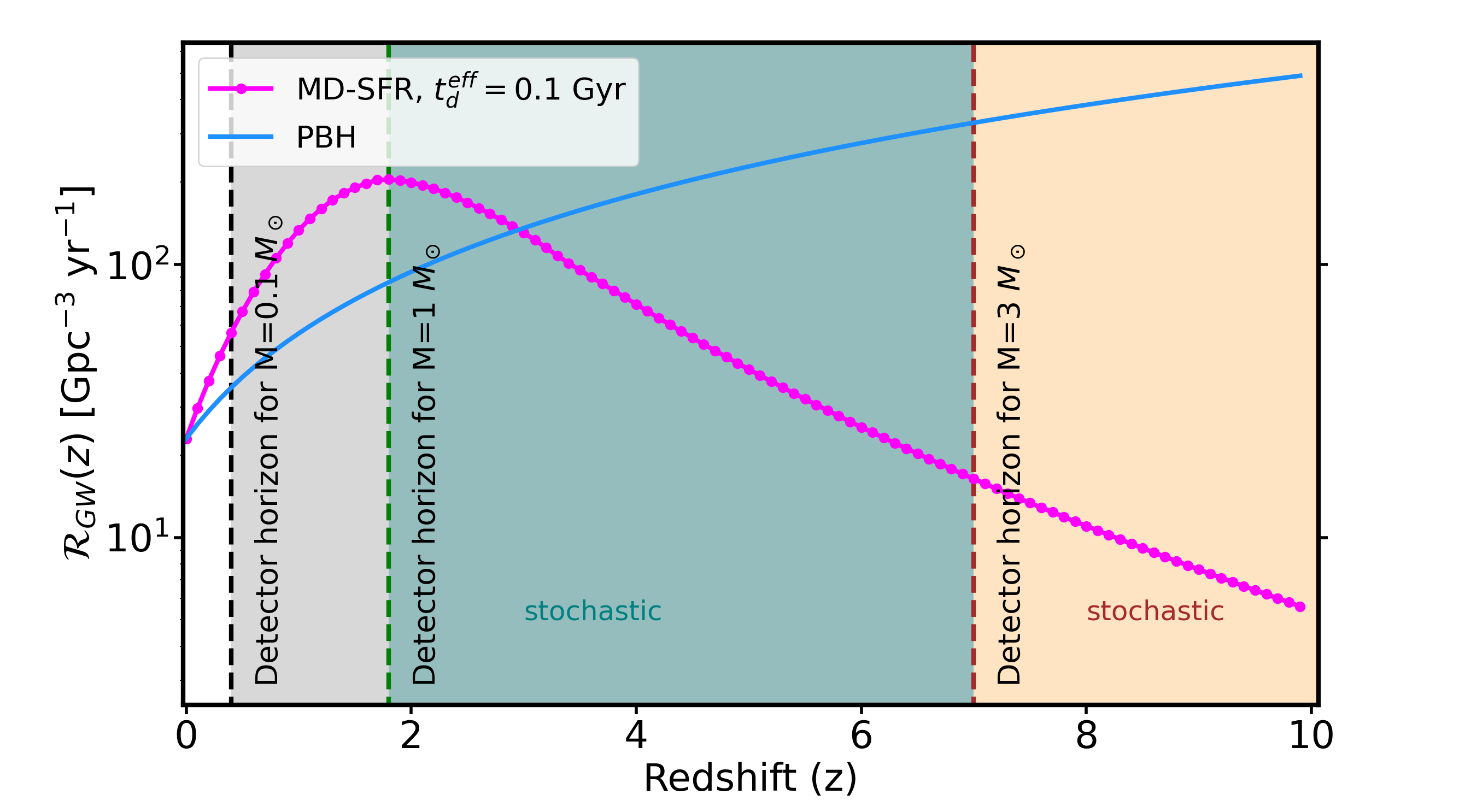}
    \caption{The GW merger rate for ABH (with $t_d^{eff}=0.1$ Gyr) and PBHs. The detector horizon (up to which one can detect an individual event) of CE for three different total source-frame masses are shown, beyond which there can be contribution to the SGWB.}
    \label{fig:merger_rate}
\end{figure}

\section{Signal from ABHs} 
The formation mechanism for BBHs involving ABHs is not yet completely certain, however, there exist several models for how such a system can occur. One example is the evolution of an isolated binary within a common envelope undergoing mass transfer (\citep{Belczynski_2016,Stevenson_2017,Abbott_2019}). The second main channel thought to form BBHs is that of dynamical mergers within densely populated globular clusters (\citep{Rodriguez_2016, Chatterjee_2017}). The observation of GWs from such mergers provides an excellent way to test the formation channels. \\
The BBH merger rate for the astrophysical sources can be written in terms of the SFR and time-delay model as 
\begin{equation}\label{eq:Rabh}
    {R}^{ABH}_{GW}(z_m)= \mathcal{N}\int^{\infty}_{z_m} dz \frac{dt_f}{dz}P(t^{eff}_d) R_{SFR}(z),
\end{equation}
where $\mathcal{N}$ is a normalisation constant such that ${R}^{ABH}_{GW}(0)$ matches the low redshift constraints possible from individual events \citep{Abbott:2018oah, Authors:2019qbw, Phukon:2021cus}. $P(t^{eff}_{d})$ is our probabilistic model describing the probability of a BBH system undergoing a merger at $z = z_{m}$ given the system was formed at $z = z_{f}$ with a time delay factor of $t_{d}^{eff}$ between the formation of stars and merging of the compact objects. We consider a logarithmic model, where $P(t^{eff}_{d}) = 1/t^{eff}_{d}$ for when $t^{eff}_{d} > t^{min}_{d}$ \citep{Mukherjee:2021ags}. 
For the star-formation rate (SFR), we take a Madau-Dickenson SFR \citep{Madau:2014bja} $R_{S F R}(z) \propto   \frac{0.015 (1+z)^{2.7}}{1+\left(\frac{(1+z)}{2.9}\right)^{5.6}}$.
The mass distribution of ABHs is considered to be a  power-law model $m^{-\beta}$ with the value of $\beta=2.35$ for the heavier BH and $1/m$ on the lighter BHs. For the BNS systems, we consider a flat distribution over the mass range $1-2\, M_\odot$. 

\section{Signal from PBHs}
Several theoretical studies have shown  possible ways to form PBHs in the early Universe \citep{1967SvA....10..602Z,1971MNRAS.152...75H,1975ApJ...201....1C, 1980PhLB...97..383K, 1985MNRAS.215..575K,Carr:2005zd, Clesse:2016vqa, Sasaki:2018dmp, Sasaki:2016jop, Ali-Haimoud:2017rtz,Raidal:2017mfl, Raidal:2018bbj, Vaskonen:2019jpv, Gow:2019pok,Jedamzik:2020ypm,Jedamzik:2020omx,DeLuca:2020jug, Atal:2020igj, DeLuca:2020qqa}. Sub-solar mass PBHs are proposed to exist theoretically in various scenarios \citep{Nakamura:1997sm, Ioka:1998nz,Carr:2019kxo,DeLuca:2020sae}. From current observations using the LIGO/Virgo GW data, constraints are achieved on sub-solar GW sources by searching for GW events from the very low redshift Universe  \citep{Abbott:2018oah, Authors:2019qbw, Phukon:2021cus}. Along with the search for individual sources from the low redshift Universe,  the SGWB is a complementary avenue to search for sub-solar GW sources which are merging at high redshift and are beyond the horizons of the current detectors.

The merger rate of the sub-solar GW sources is model-dependent, and depends on  different formation scenarios \citep{1967SvA....10..602Z,1971MNRAS.152...75H,1975ApJ...201....1C, 1980PhLB...97..383K, 1985MNRAS.215..575K,Carr:2005zd, Clesse:2016vqa, Sasaki:2018dmp, Sasaki:2016jop, Ali-Haimoud:2017rtz,Raidal:2017mfl, Raidal:2018bbj, Vaskonen:2019jpv, Gow:2019pok,Jedamzik:2020ypm,Jedamzik:2020omx,DeLuca:2020jug, Atal:2020igj, DeLuca:2020qqa}. The merger rate for the scenario in which the spatial distribution of the PBHs is Poissonian (without clustering)

\begin{align}
\begin{split}\label{mergerpbh}
    \frac{R_{PBH}(t)}{\text{Gpc$^{-3}$ yr$^{-1}$}}= &1.6\times 10^6f_{\text{sup}}f_{\text{PBH}}^{53/37} \eta^{-34/37}\bigg(\frac{M}{M_\odot}\bigg)^{-32/37}\\ & \times \bigg(\frac{t}{t_0}\bigg)^{-34/37}P_{PBH}(m_1)P_{PBH}(m_2),
    \end{split}
\end{align}
where $f_{\text{sup}}$ is the suppression factor which varies by orders of magnitude from $10^{-3}$ to 1 \citep{Raidal:2018bbj}, $f_{\text{PBH}}$ is the fraction of PBHs in dark matter, $\eta$ denotes the reduced mass ratio, $M$ denotes the total mass, $t$ denotes the proper time, and $t_0$ denotes the age of the Universe. Also, if there is spatial clustering, then the local merger rate of the PBHs can be significantly suppressed and the time dependence is going to be $(t/t_0)^{-1}$ along with a slightly different mass dependence \citep{Raidal:2017mfl, Raidal:2018bbj, Young:2019gfc, Vaskonen:2019jpv, Atal:2020igj, DeLuca:2020qqa}. 
Though the mass-dependence for the Poisson scenario and the clustering scenario is not very easy to capture from the SGWB, the redshift evolution of the merger rate for the Poisson scenario and the clustering scenario is different and can be used as a key feature to distinguish between these two scenarios. Hence we consider a model-independent parameterization of the merger rate that can be efficiently searched for different models using the SGWB data. We model the PBH merger rate as \citep{Mukherjee:2021ags}
\begin{equation}\label{pbhsfr}
R_{PBH}(z)= R_{PBH}(0, m_1,m_2)(1+z)^\alpha,
\end{equation}
where $R_{PBH}(0, m_1,m_2)$ denotes the mass-dependent merger rate at $z=0$ (which depends on $f_{PBH}$, $f_{sup}$ and PBH masses by Eq. \eqref{mergerpbh}), and $\alpha$ denotes the power-law index of the model. This parametric form will allow us to test a vast range of models from the SGWB data in an efficient way. The fiducial value of $\alpha \sim 1.28$ captures a broad class of PBH merger rates for the Poisson scenario \citep{Raidal:2017mfl, Raidal:2018bbj} and $\alpha \sim 1.44$ for the clustering scenario. For the mass distribution of the PBHs, we will consider a log-normal distribution with a characteristic mass $M_c$ and standard deviation $\sigma$, which can be expressed as \citep{Raidal:2017mfl, Raidal:2018bbj, Young:2019gfc, Vaskonen:2019jpv, Atal:2020igj, DeLuca:2020qqa} 
\begin{equation}\label{pbhmass}
    P_{PBH}(m)= \frac{1}{\sqrt{2\pi}\sigma m}\exp\bigg({\frac{-\log^2{(m/M_c)}}{2\sigma^2}}\bigg),
\end{equation}
which is motivated by  the power spectrum of small-scale density fluctuations \citep{Dolgov:1992pu,Carr:2017jsz}. Along with the contribution from coalescing binaries, the formation of PBHs also leaves signatures on the low-frequency GW  \citep{Kohri:2018awv,Espinosa:2018eve, Wang:2019kaf}.

\section{Theoretical estimation of the SGWB signal}
Using the analytical formula of the SGWB signal given in Eq. \ref{sgwb-1}, we calculate the signal for PBHs using the merger rate given in Eq. \ref{pbhsfr} for different values of the parameters $\alpha$ and $R_{GW}$. We consider merger rates as $23$ Gpc$^{-3}$ yr$^{-1}$ at redshift $z=0$ which is consistent with  observations from O1+O2 \citep{Abbott_2019}. The latest constraints on the sub-solar merger rate range from a few $\times 10^5$ Gpc$^{-3}\, \text{yr}^{-1}$ to a few $\times 10^3$ Gpc$^{-3}\, \text{yr}^{-1}$ for chirp masses in the range $0.1-1$ $M_\odot$. The mass distribution of the PBHs is taken to be  log-normal  as shown in Eq. \ref{pbhmass}.  {We consider two scenarios of PBH log-normal mass populations with $M_c=1\, M_\odot$ and $M_c=0.05\, M_\odot$, and the corresponding standard deviations $\sigma=0.1$ and $\sigma=0.01$ respectively. We also set $f_{PBH}=1$ for this plot, which in combination with the merger rate, fix the value of $f_{sup}$. For a lower value of $f_{PBH}$, the value of the SGWB will decrease according to the Eq. \eqref{mergerpbh}.} We also calculate the expected SGWB signal from the astrophysical sources for masses in the range $1\,$ $M_\odot$ to $50\,$ $M_\odot$ with the merger rate distribution according to the Madau-Dickinson SFR \citep{Madau:2014bja} with different values of the minimum value of the time delay parameter $t_d^{eff}$ Gyr with probability distribution $1/t^{eff}_{d}$, and local merger rate consistent with GWTC-2 \citep{Abbott:2020niy}.  We show the expected SGWB for a few cases in Fig. \ref{fig:sgwb}. The astrophysical contribution to the SGWB signal arises mainly from the unresolved BNS and NS-BH systems.  {The current bounds on the PBH as dark matter $f_{PBH}<6\%$ at 90\% confidence limit from the LIGO-Virgo-KAGRA data does not take into account the factor $f_{sup}$ and mentions that the bounds on the fraction of PBH in dark matter can be less constraining for $f_{sup}<1$ \cite{LIGOScientific:2021job}. There are also independent bounds on the fraction of dark matter $f_{PBH}<10^{-6}$ \cite{Chen:2019xse} in the sub-solar mass range $0.002-0.7$ M$_\odot$ (assuming a monochromatic mass spectrum) due to the absence of any signal from 11 years of NANOGrav data sets \cite{NANOGrav:2017wvv}. These bounds will be less stringent for a non-monochromatic mass spectrum considered in their analysis \cite{Chen:2019xse}. The bounds possible from the SGWB due to the mergers of the PBHs from CE and ET can provide an independent bound on the PBH fraction as dark matter in the sub-solar mass range for a different mass spectrum than considered in \cite{Chen:2019xse}. However, with the availability of both low frequency SGWB data from NANOGrav data and high frequency SGWB data from CE and ET, stringent bounds can be obtained for different mass distribution of the BBHs.}   
\begin{figure}
     \centering
         \centering
         \includegraphics[width=0.9\linewidth]{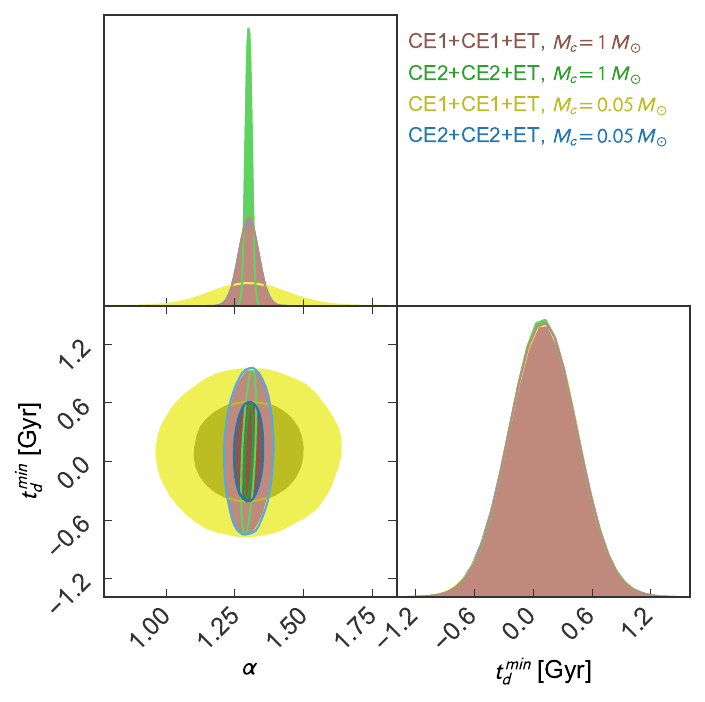}
         \caption{Fisher estimate is  obtained for $\alpha=1.28$ and a fixed value of $f_{PBH}=1$ with a $30\%$ prior on $t_d^{min}$ and for a fixed value of $f_{sup}= 2\times 10^{-3}$.}
         \label{fig:fisher_alpha}
         \end{figure}
     \begin{figure}
         \centering
         \includegraphics[width=\linewidth]{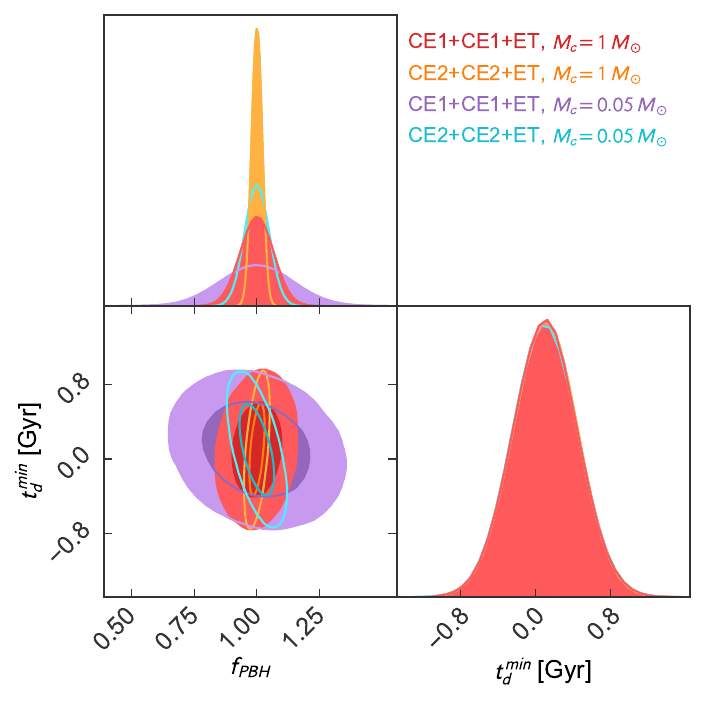}
         \caption{Fisher estimate is  obtained for $f_{PBH}=1$ for a fixed value of $\alpha=1.28$ with a $30\%$ prior on $t_d^{min}$ and for a fixed value of $f_{sup}= 2\times 10^{-3}$.}
         \label{fig:fisher_fpbh}
        \label{fig:fisher_all}
\end{figure}

\section{Fisher forecast}
The negative log-likelihood $\mathcal{L} (\hat \Omega_{gw}| \mathbf{\Theta})$ is \citep{Allen:1996vm,Allen:1997ad}
 \begin{align}\label{sgwb-len-3}
    \begin{split}
    \mathcal{L} (\hat \Omega_{gw}| \mathbf{\Theta})=& \frac{1}{2}\int_{0}^{T_{obs}} dt \int_{-\infty}^{\infty} df  \bigg(\frac{3H_0^2}{10\pi^2}\bigg)^2\\ & \times \sum_{IJ, I> J}\frac{\gamma^2_{IJ}(\hat \Omega_{IJ}(f) - \Omega_{gw} (f, \{\mathbf{\Theta}\}))^2}{f^6S_{n_I}(f)S_{n_J}(f)},
                           \end{split}
\end{align}
where observation time is denoted by $T_{obs}$, the summation over the indices (I, J) captures the combinations of different pairs of gravitational wave detector, the theoretical power spectrum of the SGWB for the parameters $\{\mathbf{\Theta}\}$ 
is shown by $\Omega_{gw} (f, \{\mathbf{\Theta}\})$ , which includes the contribution from both astrophysical and primordial sources. The detector noise power spectrum for detector $I$ (or $J$) is shown by $S_{n_I}(f)$ (or $S_{n_J}(f)$), and the normalised overlap reduction function is denoted by $\gamma_{IJ}$.
The overlap reduction function depends on the position and orientation of the detectors \citep{PhysRevD.46.5250, PhysRevD.48.2389, Allen:1997ad}. For the third generation detectors, we consider the current sites of the GW detectors \citep{TheLIGOScientific:2014jea} for two Cosmic Explorer detectors and one Einstein telescope. 

We perform a Fisher analysis \citep{1934RSPSA.146....1F, 1950ArM.....1..141K, Tegmark:1996bz} to access the possibility of measuring the sub-solar SGWB with the third generation GW detectors, and the corresponding implications for  distinguishing between compact objects of astrophysical and primordial origin. The Fisher estimation of an error on the parameters $\theta_i$ can be obtained as $\sigma^F_{\theta_i}= \sqrt{\{\mathbf{F}^{-1}\}_{ii}}$, where $\mathbf{F}_{ij} \equiv \langle \frac{\partial^2 \mathcal{L}}{\partial \theta_i\partial \theta_j}\rangle$ is the Fisher matrix for $\{ij\}$ components. 
The 
Fisher error provides the Cramer-Rao bound indicating the minimum error ($\sigma_i \geq \sigma_i^F$)  achievable from the third generation GW detectors. In future work, this method will be  extended into a completely Bayesian analysis following our previous study \citep{Mukherjee:2021ags}. 

Assuming the location of the two CE detectors at the current location of LIGO-Hanford and LIGO-Livingston, and Einstein Telescope (ET) at the site of Virgo (similar to previous studies \citep{Sachdev:2020bkk, Martinovic:2020hru}), we explore the possibility of measuring the PBH population of sub-solar masses and the capability of distinguishing it from the astrophysical sources using the SGWB signal.  {The strength of the SGWB due to astrophysical sources depends on the BH  merger rates, which in turn depend on the delay time distribution between the formation of a star and merging of compact objects (as shown in Fig. \ref{fig:sgwb}). We show using a Fisher analysis how we can distinguish between the delay time distribution (related to the astrophysical sources) and PBH merger rate and its fraction from the SGWB detectable from the third generation detectors.}

We show a Fisher forecast for both these scenarios with a pessimistic value of the suppression factor $f_{sup}= 2\times 10^{-3}$ \citep{Raidal:2018bbj, 2021JCAP...03..068H} using a network of two CE detectors and one ET detector in Fig. \ref{fig:fisher_alpha} and \ref{fig:fisher_fpbh} respectively for the parameter $\alpha$ and $f_{PBH}$, integrated over an observation time of five years using the mass distribution discussed in the previous section. For these results, we have a $30\%$ prior on the value of the total local merger rate  and on the time-delay parameter $t^{eff}_{d}$ (shown in Fig. \ref{fig:fisher_alpha} and \ref{fig:fisher_fpbh}) which should be possible to infer from the individual well-detected astrophysical sources \citep{Vitale:2018yhm}. We show that using a network of two CE detectors and one ET, we can obtain a limit on either $\alpha$ or $f_{PBH}$ with about $1$-$10\%$ precision, if the suppression factor $f_{sup}$ is minimal and for the values of $M_c= 1-0.05\, M_\odot$. However if the suppression factor $f_{sup}=1$, then one can constrain $f_{PBH}$ to a level of $  \sim 0.001\%.$
This method will make it possible to discover (or constrain) sub-solar PBHs as dark matter candidates from the third generation GW detectors, even if the suppression factor $f_{sup}$ is minimal, as is found in recent studies \citep{Raidal:2018bbj, 2021JCAP...03..068H}, and distinguish the contribution from ABH sources such as BNSs and BBHs from very high redshift sources. Moreover, the measurement of the value of $\alpha$ with $1-10\%$ accuracy will make it possible to also distinguish between the Poisson and the clustering scenarios of PBHs. As a result, we should be able to shed light on the formation channel of PBHs. Further improvements are possible if a stricter prior on $t_d^{eff}$ is attainable from the individual detected events \citep{Vitale:2018yhm} including  lensed events \citep{2013JCAP...10..022P, 2014JCAP...10..080B, 2015JCAP...12..006D,Mukherjee:2021qam}. 

If no priors are implemented on the local merger rate and the time-delay parameter, then it is not possible to break the degeneracy between these parameters and the parameters related to a PBH population such as the power-law index of PBH merger rate $\alpha$ and the fraction of PBHs in the form of dark matter $f_{PBH}$. This arises because the SGWB signal can be detected with a high signal-to-noise ratio (SNR) for only a limited frequency window (i.e. for the frequency values $f<f_{char}= c/2D$ \citep{PhysRevD.46.5250, PhysRevD.48.2389, Allen:1997ad}\footnote{$D$ denotes the distance between the two GW detectors.}). As a result, over this limited frequency range, the shape of the power spectrum of the SGWB signal originating from the BNS or sub-solar BHs looks very similar, and cannot be used to break the degeneracy between these parameters, unlike the case for heavier PBHs with $M_c= 30\, M_\odot$ which exhibit a different shape over this accessible frequency range \citep{Mukherjee:2021ags}. For sub-solar PBH searches, the only useful information that can be extracted from the SGWB signal is the amplitude of the power spectrum. As a result, the amplitude of the SGWB power spectrum can be used to extract only one parameter, either the power-law form $\alpha$ (assuming a fixed value of PBH fraction over dark matter $f_{PBH}$) or the $f_{PBH}$ parameter. One can also explore the characteristic mass-scale $M_c$, if the other parameters ($\alpha$, $f_{PBH}$, $f_{sup}$) are kept fixed. Further improvements are feasible if individual detections of sub-solar events are possible with CE and ET. While this paper was in preparation, papers appeared \citep{DeLuca:2021hde,Wang:2021djr} that independently explored related prospects for detecting PBHs with future GW  detectors via the SGWB. 

\section{Conclusions} We show the prospect of discovering sub-solar PBHs as the dark matter candidate using the SGWB which is detectable from the third generation detectors, and the capability to distinguish between the Poisson and clustering scenario of PBHs. One of the key signatures to distinguish ABHs from PBHs is through its mass distribution and the redshift evolution of the GW merger rate. These sources are only possible to be detected as individual events from the very local Universe with the current generation detectors, and also up to relatively low redshift from the third-generation detectors. But it can be detected using the SGWB arising from coalescing binaries at high redshift. 
We show that the third generation detection will be able to distinguish between astrophysical sources and PBHs and we will be able to hunt for lighter PBHs ($< 1.4\, M_\odot$) as dark matter. We can measure $f_{PBH}$ or the $\alpha$ parameter with sub-percent accuracy, even if the PBH merger rate is highly suppressed. This will shed light on whether PBHs are spatially clustered or follow a Poisson distribution. In  future work, we will explore the time dependence of the SGWB to distinguish  the ABH/PBH signal  \citep{10.1093/mnras/stz3226,Mukherjee:2020jxa} from the cosmological background \citep{Starobinsky:1979ty,Turner:1996ck, Kibble:1976sj,Kosowsky:1992rz,Kamionkowski:1993fg,Watanabe:2006qe,Damour:2004kw,Martin:2013nzq}.   

\section*{Acknowledgments}
S. M. acknowledges useful discussions and inputs from Juan Garcia-Bellido, Sebastien Clesse, and Daniele Steer on this work. A part of this work is carried out under the Master's program at the University of Amsterdam. This work is part of the Delta ITP consortium, a program of the Netherlands Organisation for Scientific Research (NWO) that is funded by the Dutch Ministry of Education, Culture, and Science (OCW).  This work has made use of the Infinity Cluster and the Horizon Cluster hosted by Institut d'Astrophysique de Paris. We thank Stephane Rouberol for smoothly running both the clusters. We acknowledge the use of following packages in this work: Astropy \citep{2013A&A...558A..33A, 2018AJ....156..123A}, Giant-Triangle-Confusogram \citep{Bocquet2016}, IPython \citep{PER-GRA:2007}, Matplotlib \citep{Hunter:2007},  NumPy \citep{2011CSE....13b..22V}, and SciPy \citep{scipy}. The authors would like to thank the  LIGO/Virgo scientific collaboration for providing the noise curves. LIGO is funded by the U.S. National Science Foundation. Virgo is funded by the French Centre National de Recherche Scientifique (CNRS), the Italian Istituto Nazionale della Fisica Nucleare (INFN), and the Dutch Nikhef, with contributions by Polish and Hungarian institutes. This material is based upon work supported by NSF’s LIGO Laboratory which is a major facility fully funded by the National Science Foundation.

 \section*{Data Availability}
The data underlying this article will be shared at request to the corresponding author. 

\bibliographystyle{mnras}
\bibliography{main_mnras}
\label{lastpage}
\end{document}